%% file: skeleton.tex
\documentclass[a4paper,11pt]{article}
\usepackage{pos}
\usepackage[separate-uncertainty=true, range-units=single,range-phrase=--]{siunitx}
\usepackage{xspace}
\usepackage{sidecap}
\usepackage{hyperref}
\usepackage{cleveref}
\usepackage{graphicx}
\usepackage{array}
\usepackage{booktabs}
\usepackage{enumitem}
\usepackage{wrapfig}
\usepackage[subrefformat=parens,labelfont=bf,font={stretch=1.15}]{subcaption}

\usepackage[
  backend=biber,
  style=phys,
  biblabel=brackets,
  articletitle=false,
  doi=true,
  eprint=true,
  url=true,
  giveninits=true,
  maxnames=2
]{biblatex}
\newcommand{\Xmax}{\ensuremath{X_{\mathrm{max}}}\xspace}
\newcolumntype{M}[1]{>{\centering\arraybackslash}m{#1}}

\newcommand{\plotwidth}{.9\textwidth}

\title{Simulating radio emission from air showers with CORSIKA 8}

\manuallySeparateAuthors
\author*[a]{Nikolaos Karastathis}
\author[b]{,   Remy Prechelt}
\author[c]{,   Juan Ammerman-Yebra}
\author[d]{,   Maximilian Reininghaus}
\author[a,e]{~and   Tim Huege}

\affiliation[a]{Institut für Astroteilchenphysik, Karlsruher Institut für Technologie, Karlsruhe, Germany}

\affiliation[b]{Department of Physics and Astronomy, University of Hawai'i M\=anoa, Honolulu, USA}

\affiliation[c]{Instituto Galego de Física de Altas Enerxías, Universidade de Santiago de Compostela, Santiago de Compostela, Spain}

\affiliation[d]{Institut für Experimentelle Teilchenphysik, Karlsruher Institut für Technologie, Karlsruhe, Germany}

\affiliation[e]{Astrophysical Institute, Vrije Universiteit Brussel, Brussels, Belgium}

\emailAdd{nikolaos.karastathis@kit.edu}

\onbehalf{for the CORSIKA 8 Collaboration} 

\abstract{CORSIKA 8 is a new framework for air shower simulations implemented in modern C++17, based on past experience with existing codes like CORSIKA 7. The flexible and modular structure of the project allows the development of independent modules that can produce a fully customizable air shower simulation. The radio module in particular is designed to treat the signal propagation and electric field calculation to each antenna in an autonomous and flexible way. It provides the possibility to simulate simultaneously the radio emission calculated with two independent time-domain formalisms, the “Endpoint formalism” as implemented in CoREAS and the “ZHS” algorithm as ported from ZHAireS. Future development for the simulation of radio emission from particle showers in complex scenarios, for example cross-media showers penetrating from air into ice, can build on the existing radio module, re-using the established interfaces.
In this work, we will present the design and implementation of the radio module in CORSIKA 8, and show a direct comparison of radio emission from air  showers simulated with CORSIKA 8, CORSIKA 7 and ZHAireS.}

\ConferenceLogo{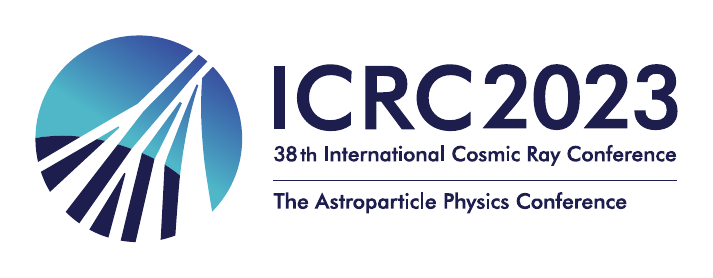}

\FullConference{%
38th International Cosmic Ray Conference (ICRC2023)\\
  26 July - 3 August, 2023\\
  Nagoya, Japan}

\begin{document}
\maketitle

\section{Introduction}
The radio detection technique of extensive air showers has experienced a remarkable renaissance over the past 20 years~\cite{Huege:2016veh} and has established itself to be a competitive technique to particle and fluorescence detection.  Reconstructing the main properties of the air shower and of the primary particle using experimental data is a highly complex procedure. To overcome this, detailed particle-level Monte Carlo simulations render themselves very useful. The community depends on existing state of the art simulation tools like CoREAS~\cite{Huege:2013vt} (implemented in CORSIKA~7~\cite{Heck:1998vt}) and ZHAireS~\cite{Alvarez-Muniz:2010hbb} to simulate the radio emission from extensive air showers at the microscopic level. Two different formalisms for calculating the radio emission from
the particle tracks are used by these tools. CoREAS uses the ``Endpoint'' formalism
\cite{James:2010vm,Ludwig:2010pf} and ZHAireS uses the ``ZHS''~\cite{Alvarez-Muniz:2010wjm} formalism. However, these implementations rest upon shower simulation codes that exist for decades and are inherently limited by them. By default, this makes the radio implementations non flexible to specific physical scenarios and deprives them of taking advantage of new computing technologies to perform simulations for the diverse array of current and future experiments. Moreover, the calculation of the radio emission is one of the most computationally demanding modules (especially for ultra high-energy showers and many antenna positions). This will make computation times unmanageable, for the proposed next-generation experiments whose radio detectors are growing significantly in terms of size and number. To tackle these constraints, we have implemented the first radio emission module for the CORSIKA~8 (C8) simulation framework~\cite{Engel:2018akg}, \cite{Karastathis:2021akf}. It is designed to be highly configurable, user-extensible and ready to quickly adapt technologies like multithreading \cite{AugustoICRC2023} to directly address the
limitations of the current simulation tools and support the next generation of
radio detection experiments.

\section{The architecture of the radio module}

C8 incorporates three main concepts in its design, which are modularity, flexibility and extensibility. The radio module follows this pattern and consists of four top level, user-configurable and swappable components. These are the \textit{track filter}, the \textit{formalism}, the \textit{propagator} and the \textit{antenna} and their functionality has been described in~\cite{Karastathis:2021akf}. A \textit{radio process} is constructed by these separate components and is added in C8's process sequence list. As part of the process sequence, the \textit{radio process} has direct access to the particle tracks and the environment of the simulation, from where it draws all the information needed to calculate the radio emission from the air shower. These components are designed to be flexible enough and have clear guidelines, so even users can easily update them to tailor the simulation to their specific experimental needs. 

At this point, it is worth discussing about the advancements we have included in the \textit{propagator} since \cite{Karastathis:2021akf}. All the propagators we have implemented use a straight-ray approximation. We include an analytic ray path solver propagator that can only be used in media with uniform or exponential refractive indices, ideal for tests and fast simulations; and an \textit{integrating propagator} that numerically integrates the time delay along each propagation path and can therefore work in arbitrarily complex media where no analytic solution exists. We have now developed a  \textit{tabulated propagator} ideal to use for vertical showers (zenith $< \ang{60}$) that approximates the atmosphere to be flat and has the advantage of small computing times due to the fact that it is initialized once, in the beginning of the simulation. This propagator uses the Gladstone-Dale law~\cite{GladstoneDale} which states that refractivity in the atmosphere is proportional to the density. Similar approximations also take place in other shower codes like CORSIKA~7 for example.

\section{Validation and improvements of the radio module}

The radio module consists of unit tests and validation tests, following C8's development guidelines, which are initiated with every update of the code. A standard module validation test we perform is the synchrotron emission of an electron gyrating in a circle in a constant magnetic field as explained in \cite{Karastathis:2021akf}. This validates not only the radio module independently of C8, but also the magnetic field tracking algorithm of C8 since the emitted pulse is calculated using a \textit{manual} tracking and C8's standard tracking algorithm and the tracks along with the 2 pulses that come out match perfectly \cite{Karastathis:2021akf}.

In \cite{Karastathis:2021akf,NikosARENA2022} comparisons of \SI{10}{\TeV} electron induced showers were presented, showing promising results. However, technical difficulties affected our results and we identified and addressed these issues successfully. We can now confirm that for non high-energy showers, C8 and radio have no problems and work as intended. Several technical advancements also took place in the radio module to increase its performance. The most notable one is the development of a version of the radio module which supports the use of multithreading \cite{AugustoICRC2023}. With this feature, the radio module can now detect available threads in the CPU and assign different antennas to different threads in order to perform the radio emission calculation in parallel. Although this version is still not merged in the main C8 code, it acts as a proof of principle that the radio module is written in a modern way, ready to adapt new technologies.

\begin{wrapfigure}{r}{8cm}
    \includegraphics[width=8cm]{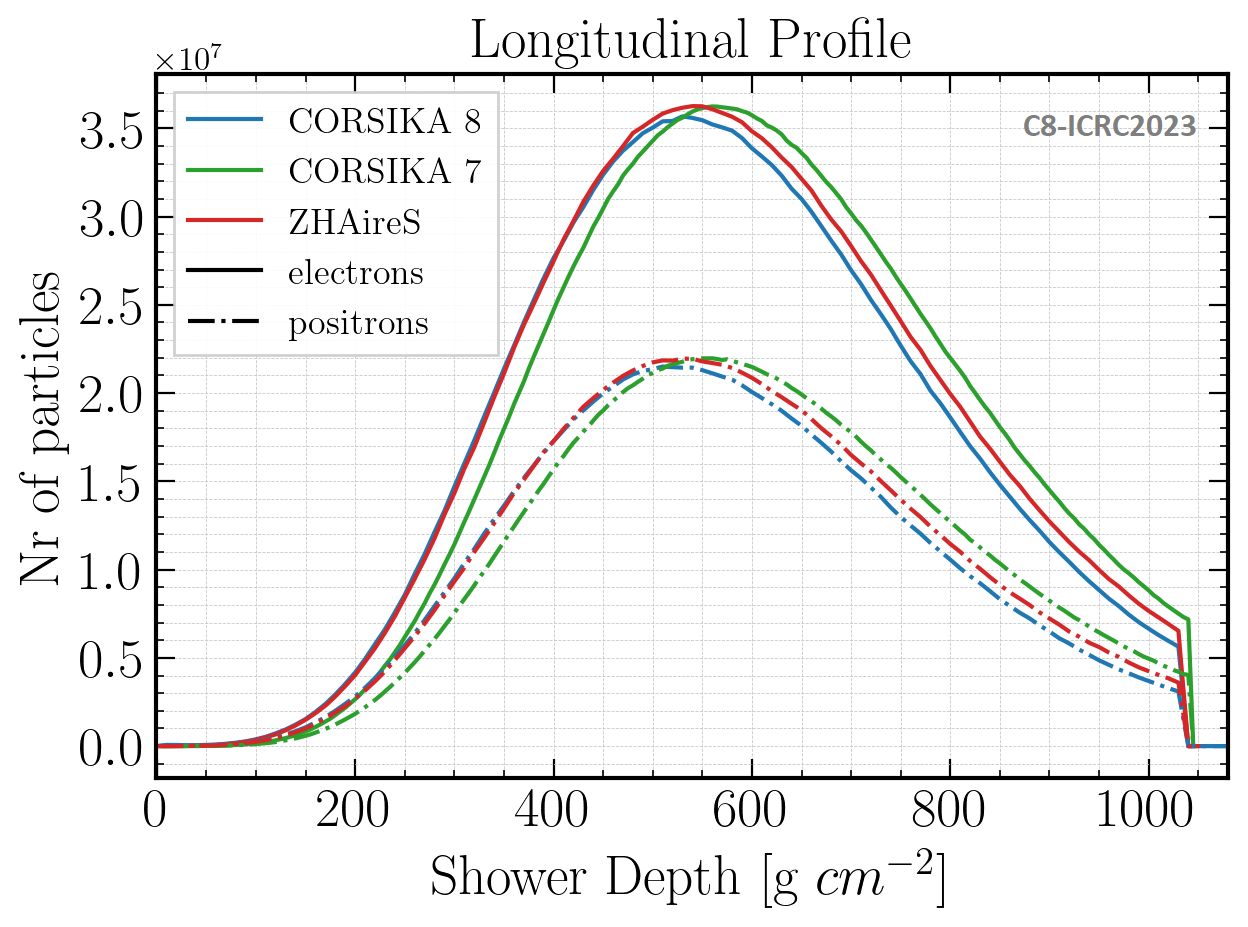}
    \caption{Longitudinal profiles of air showers simulated with C8, C7 and ZHAireS. The number of electrons and positrons with respect to grammage is shown.}\label{fig:profile}
\end{wrapfigure}

\section{Comparison of simulations of iron-induced extensive air showers}

\begin{figure}[h]
\centering
\begin{subfigure}{6cm}
    \centering
    \includegraphics[width=6cm]{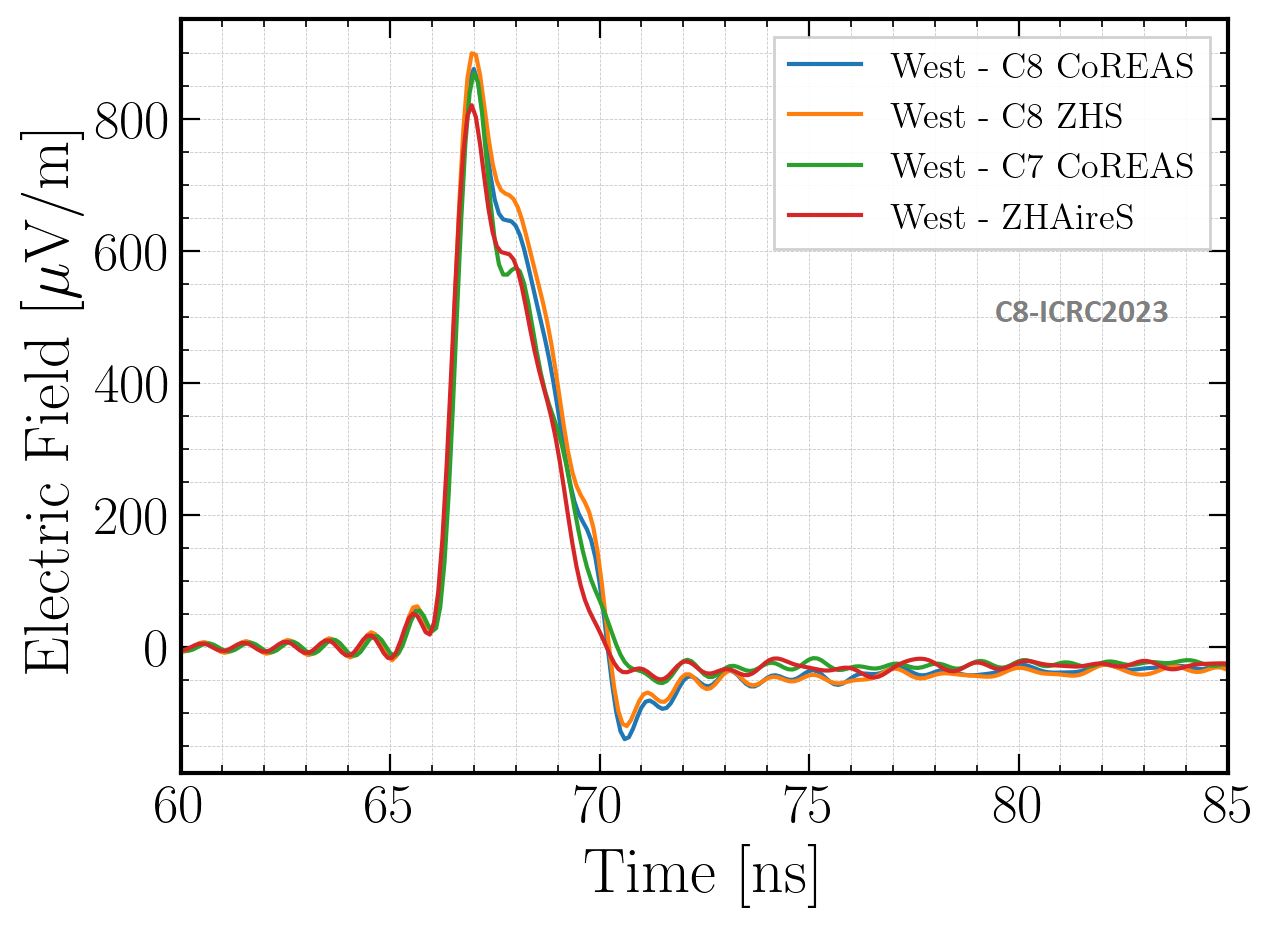}
    \subcaption[c]{\SI{50}{\metre}.}
    \label{fig:west50}
\end{subfigure} \hspace{10mm}
\begin{subfigure}{6cm}
    \centering
    \includegraphics[width=6cm]{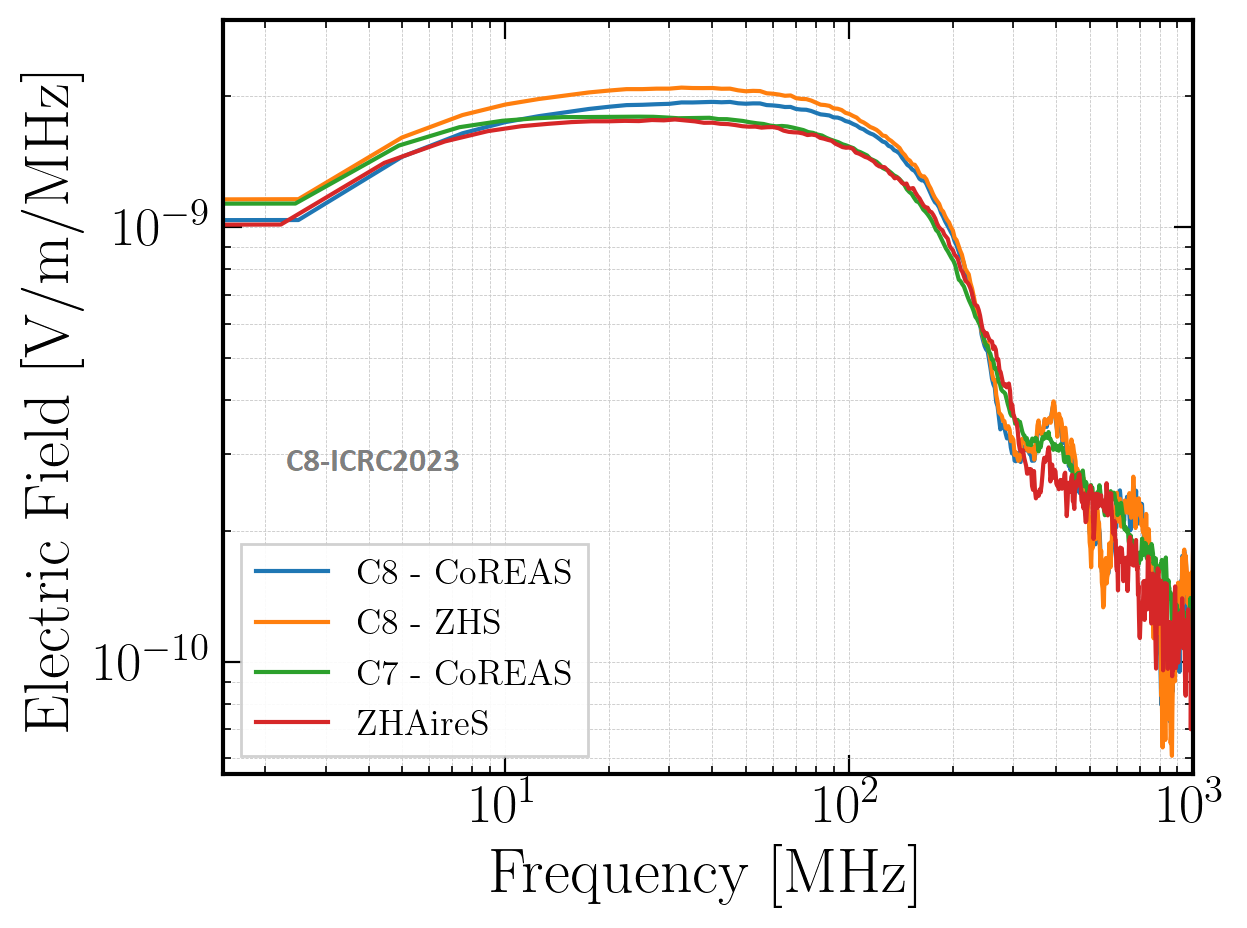}
    \subcaption[c]{\SI{50}{\metre}.}
    \label{fig:west50fft}
\end{subfigure}\\
\begin{subfigure}{6cm}
    \includegraphics[width=6cm]{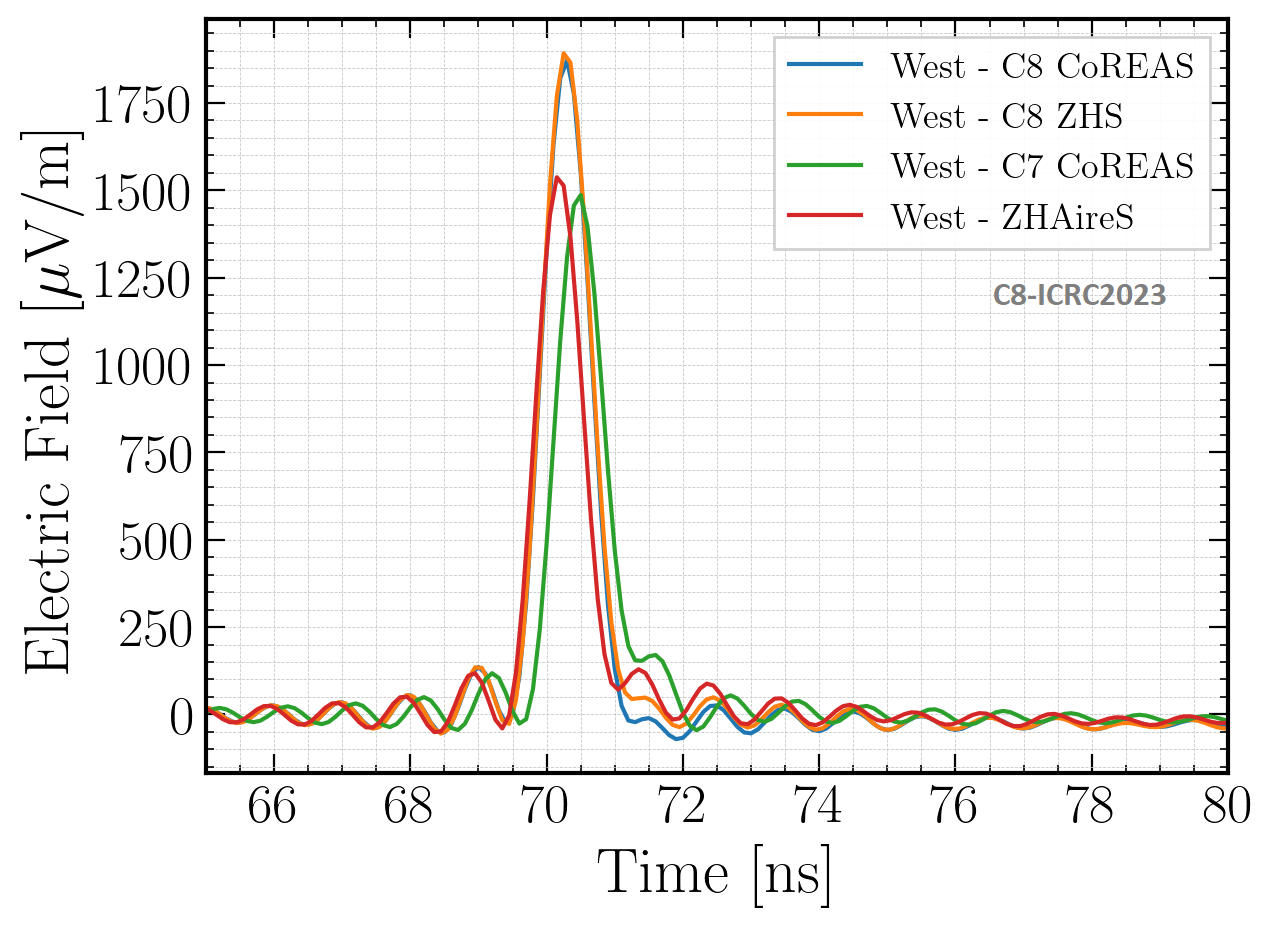}
    \subcaption[c]{\SI{100}{\metre}.}
    \label{fig:west100}
\end{subfigure} \hspace{10mm}
\begin{subfigure}{6cm}
    \includegraphics[width=6cm]{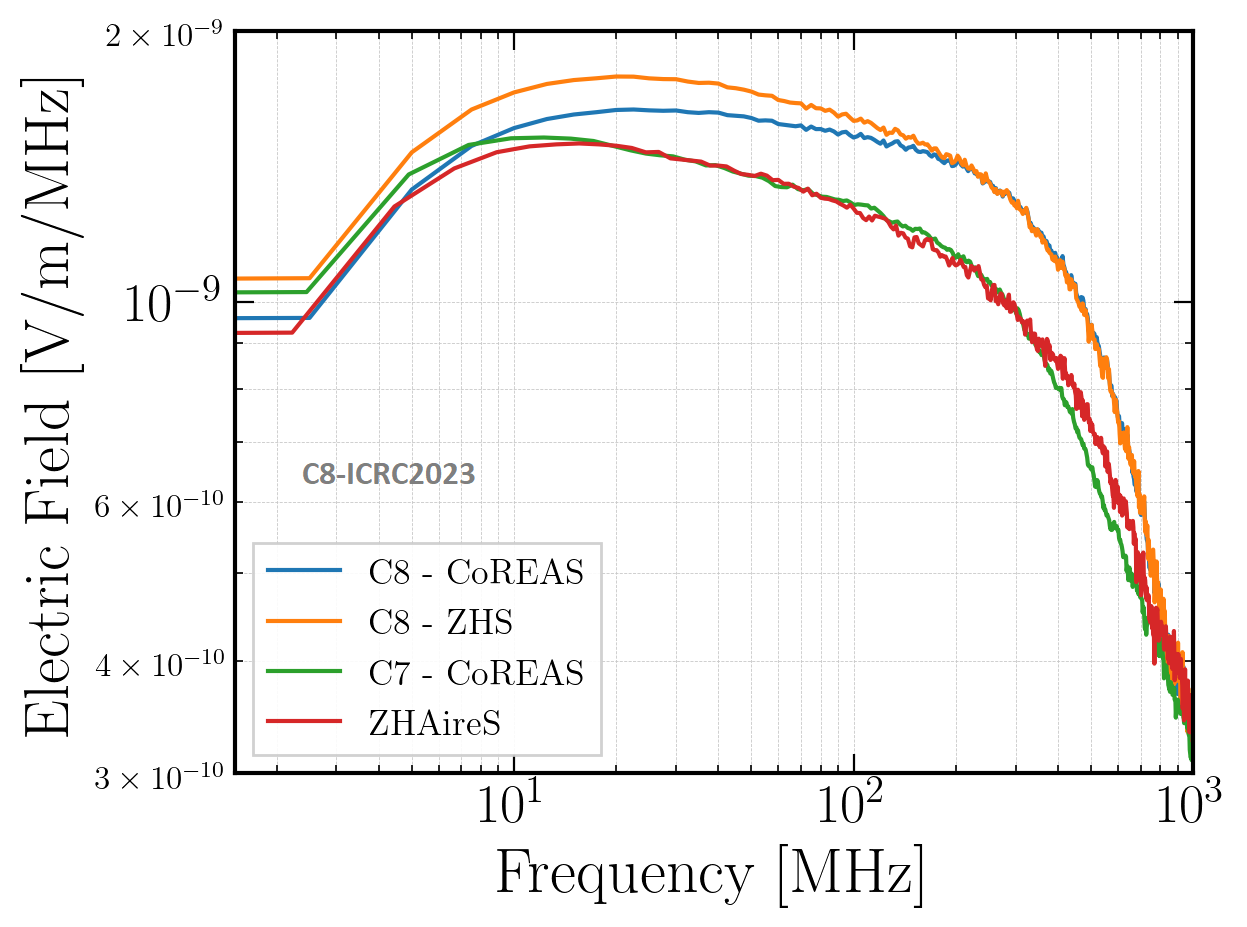}
    \subcaption[c]{\SI{100}{\metre}.}
    \label{fig:west100fft}
\end{subfigure}\\
\begin{subfigure}{6cm}
    \centering
    \includegraphics[width=6cm]{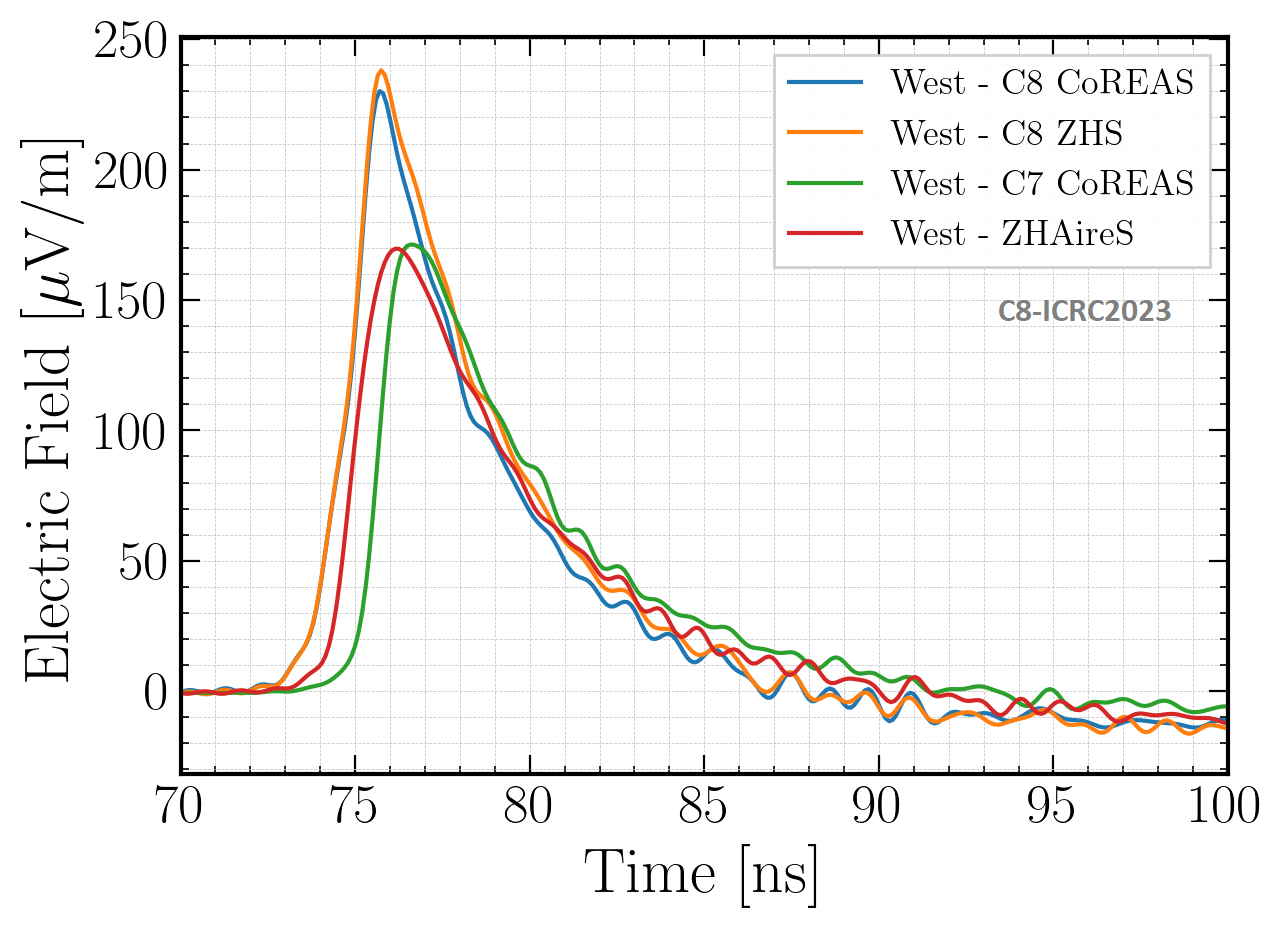}
    \subcaption[c]{\SI{200}{\metre}.}
    \label{fig:west200}
\end{subfigure} \hspace{10mm}
\begin{subfigure}{6cm}
    \centering
    \includegraphics[width=6cm]{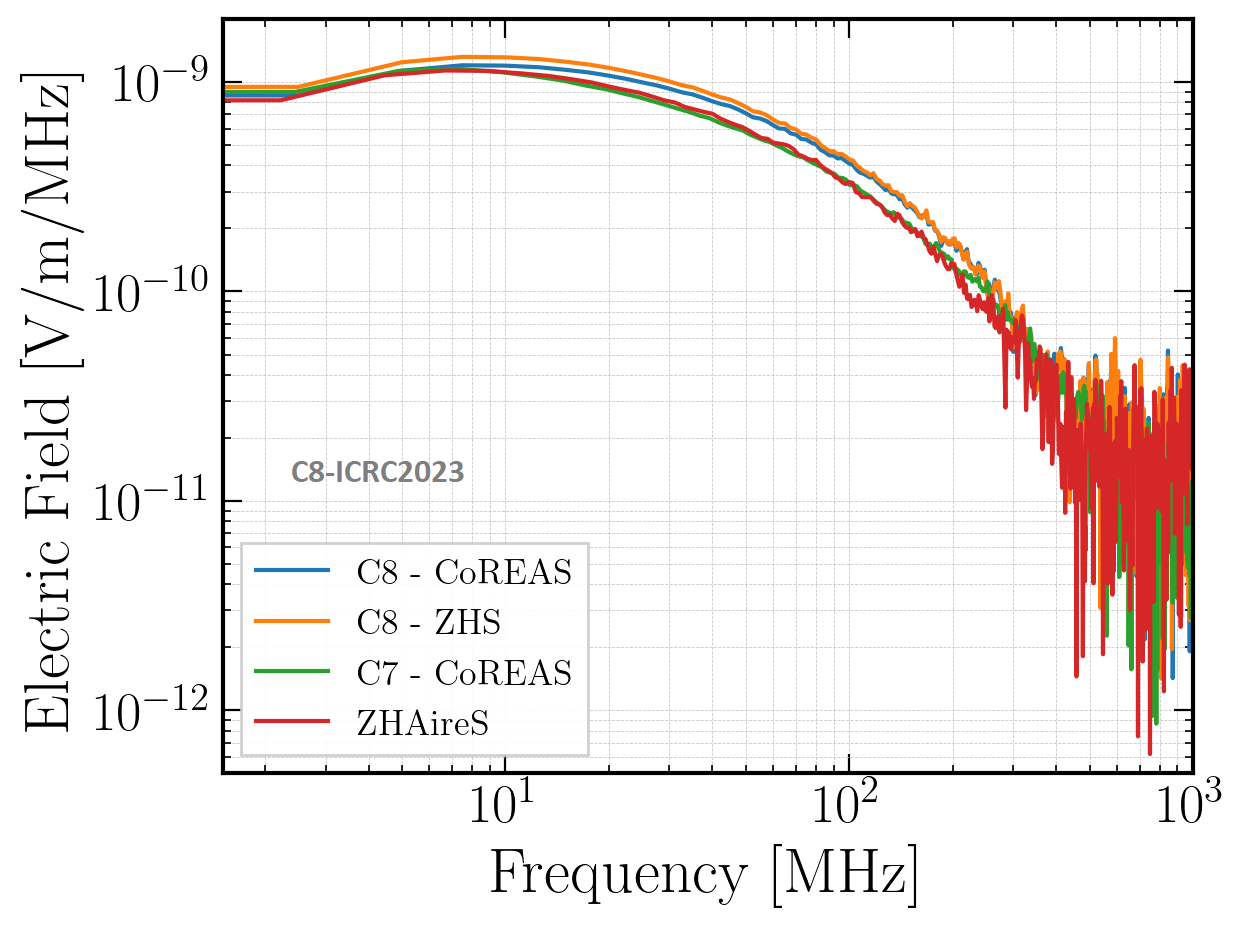}
    \subcaption[c]{\SI{200}{\metre}.}
    \label{fig:west200fft}
\end{subfigure}
\caption{Signal pulse and frequency spectra comparison in \SIrange{0}{1}{\giga\hertz} band, for various antenna distances from the shower core - Geomagnetic contribution.}
\end{figure}

\begin{figure}[]
\centering
\begin{subfigure}{6cm}
    \centering
    \includegraphics[width=6cm]{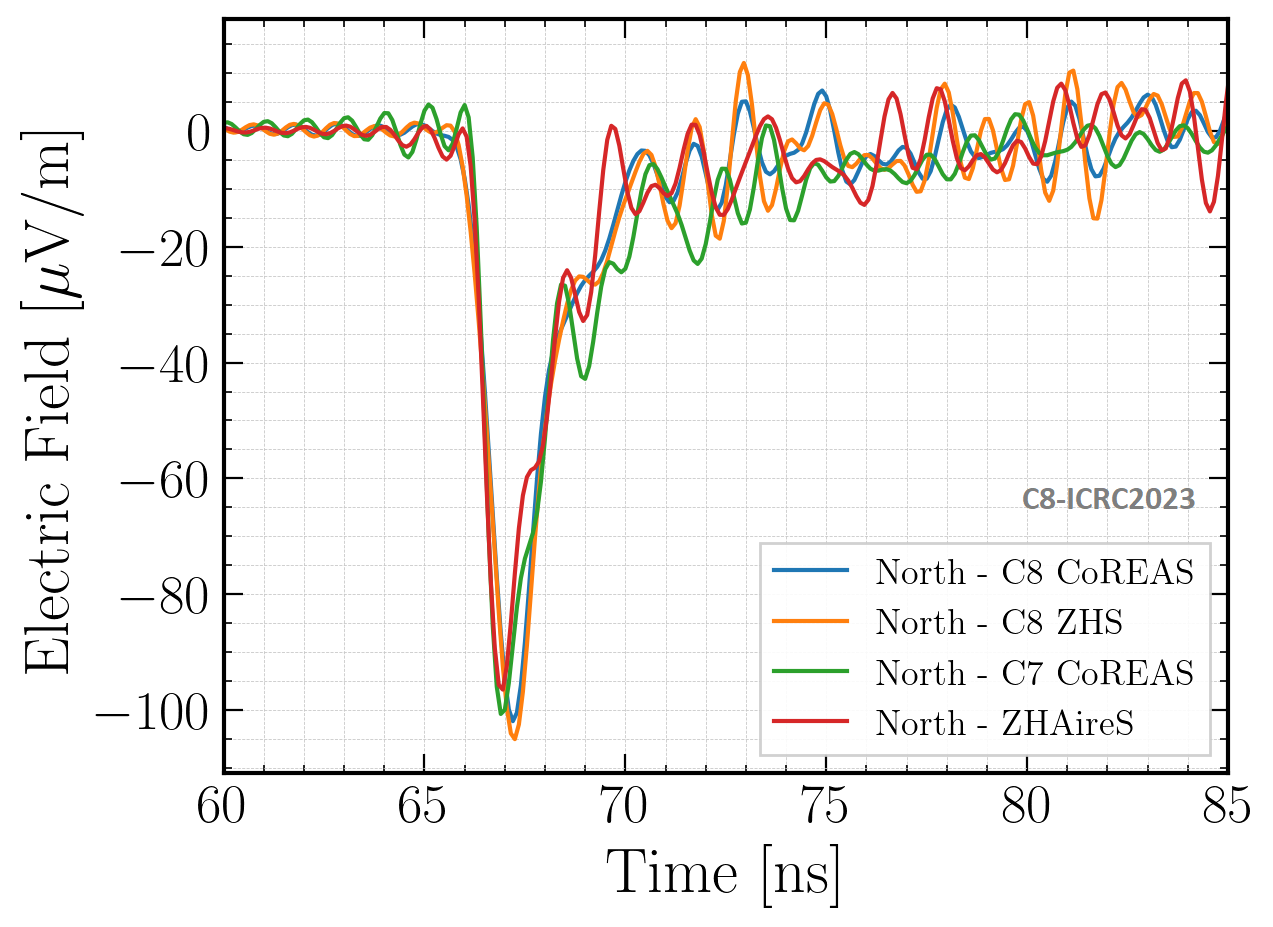}
    \subcaption[c]{\SI{50}{\metre}.}
    \label{fig:north50}
\end{subfigure} \hspace{10mm}
\begin{subfigure}{6cm}
    \centering
    \includegraphics[width=6cm]{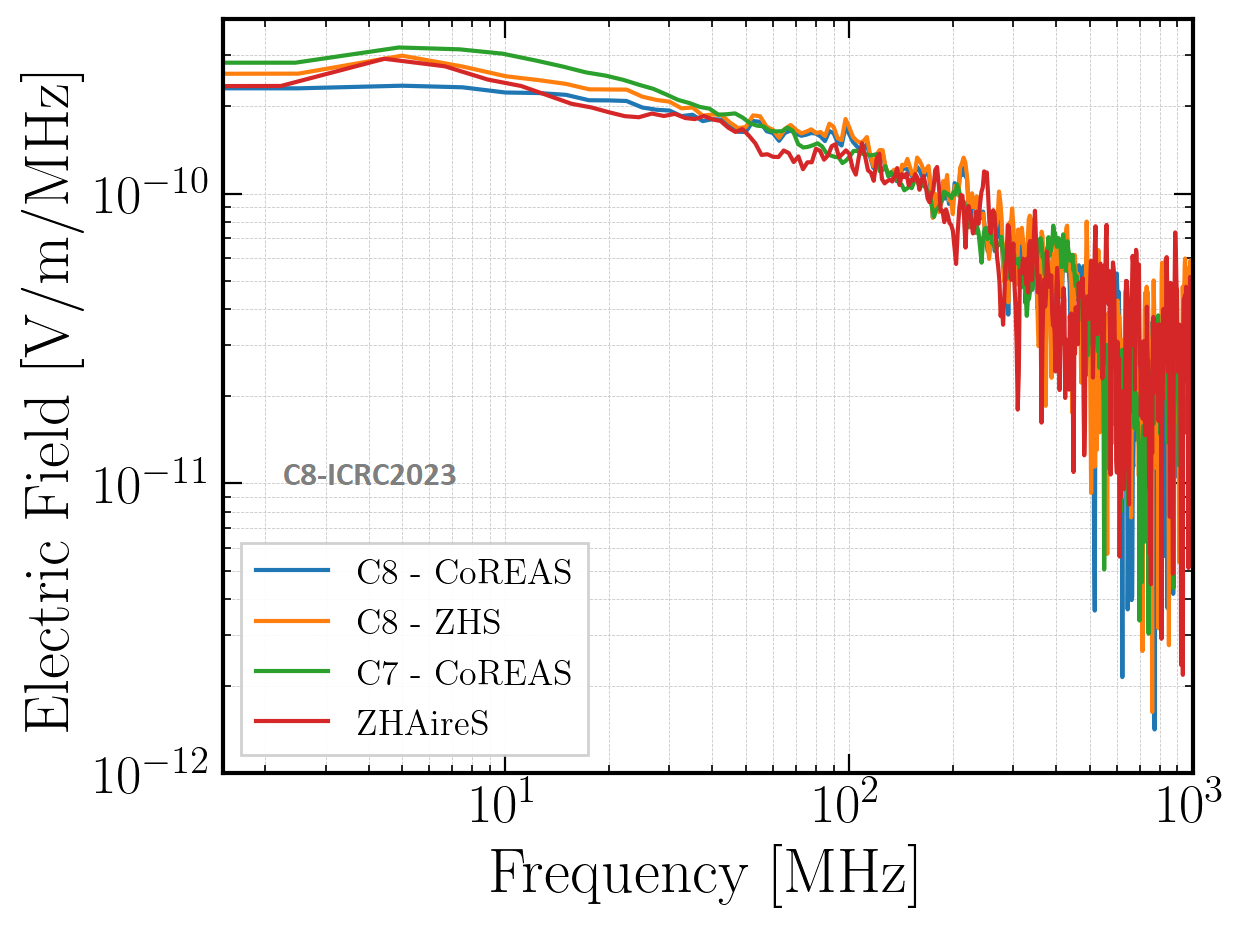}
    \subcaption[c]{\SI{50}{\metre}.}
    \label{fig:north50fft}
\end{subfigure}\\
\begin{subfigure}{6cm}
    \centering
    \includegraphics[width=6cm]{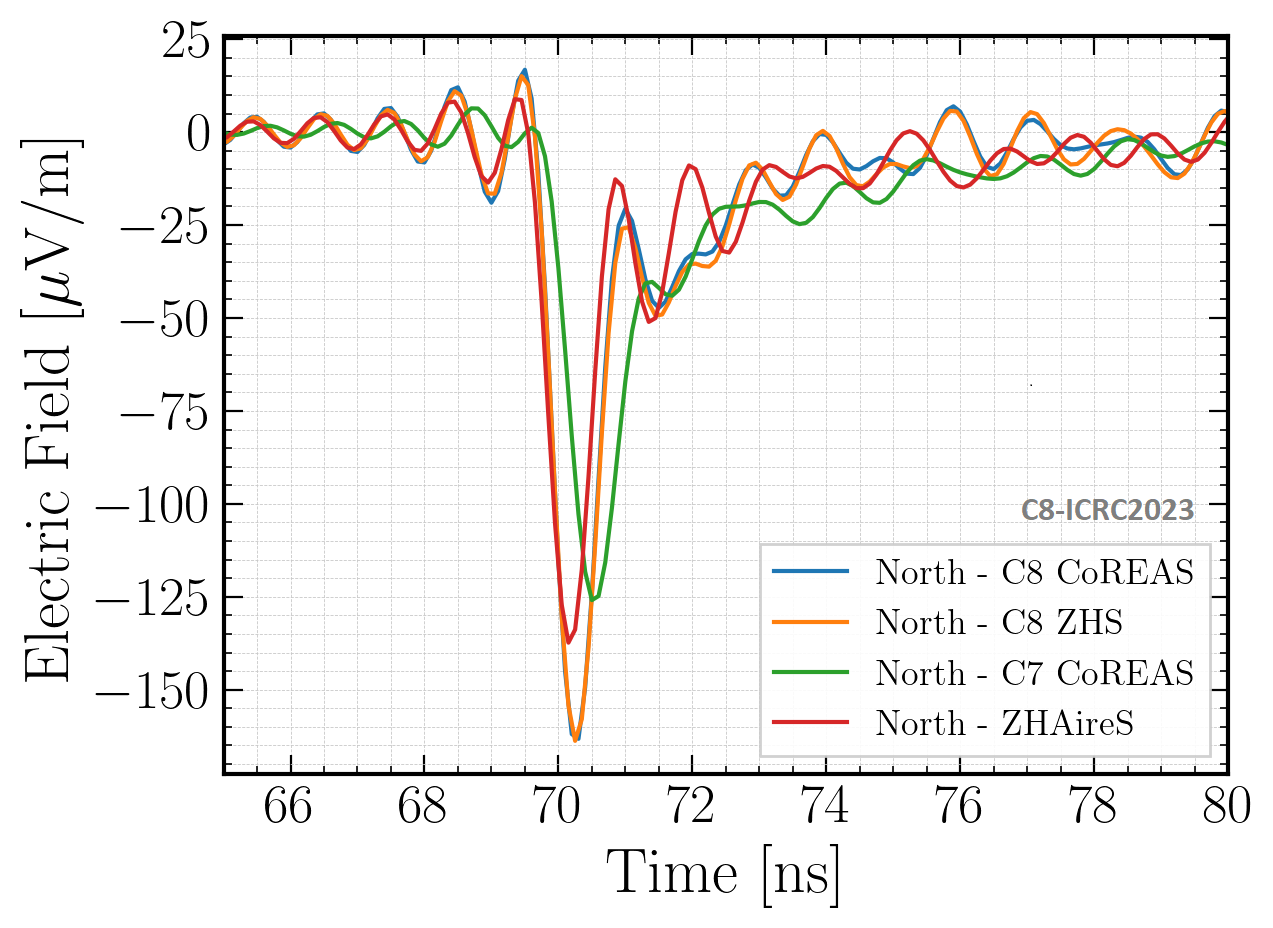}
    \subcaption[c]{\SI{100}{\metre}.}
    \label{fig:north100}
\end{subfigure} \hspace{10mm}
\begin{subfigure}{6cm}
    \centering
    \includegraphics[width=6cm]{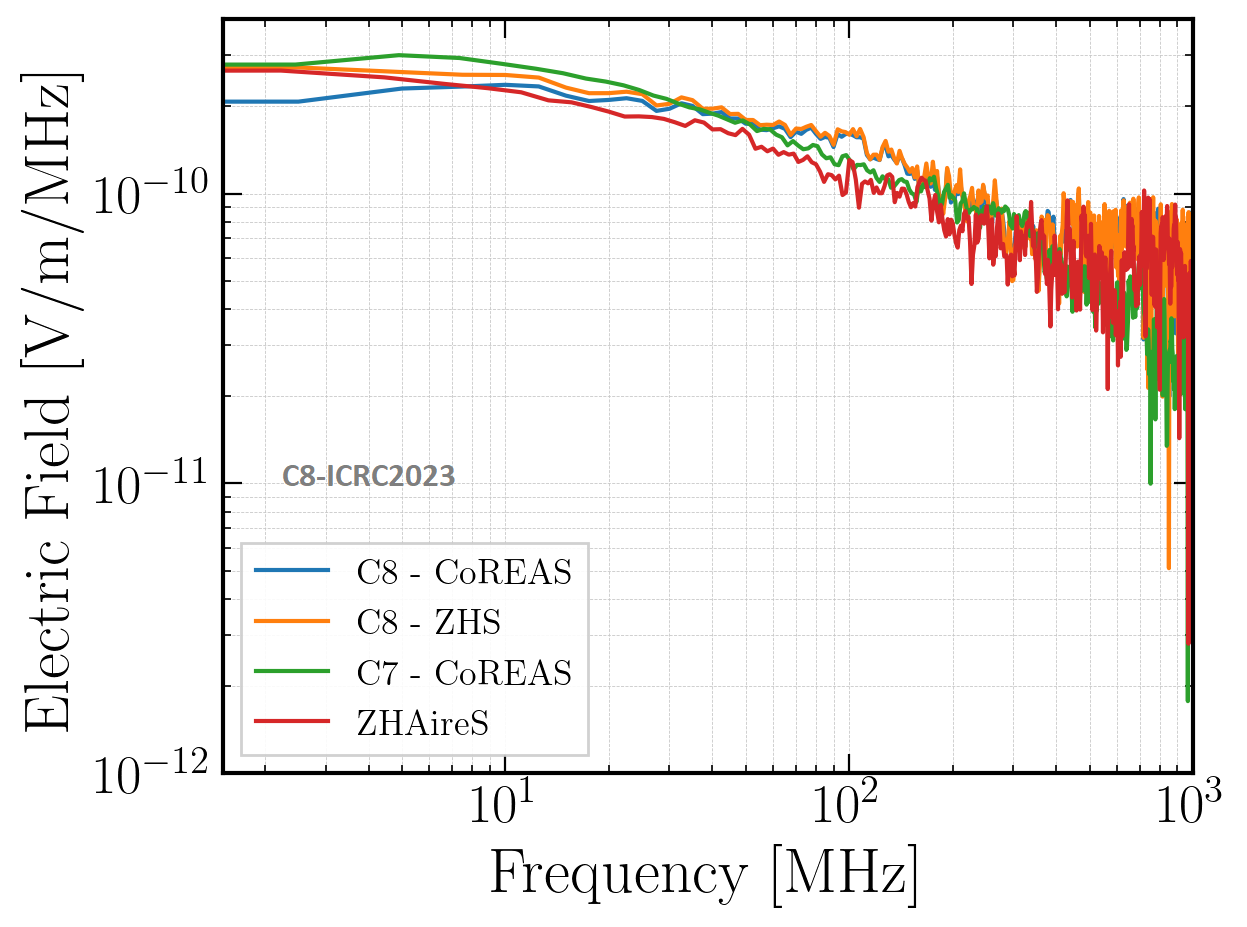}
    \subcaption[c]{\SI{100}{\metre}.}
    \label{fig:north100fft}
\end{subfigure}\\
\begin{subfigure}{6cm}
    \centering
    \includegraphics[width=6cm]{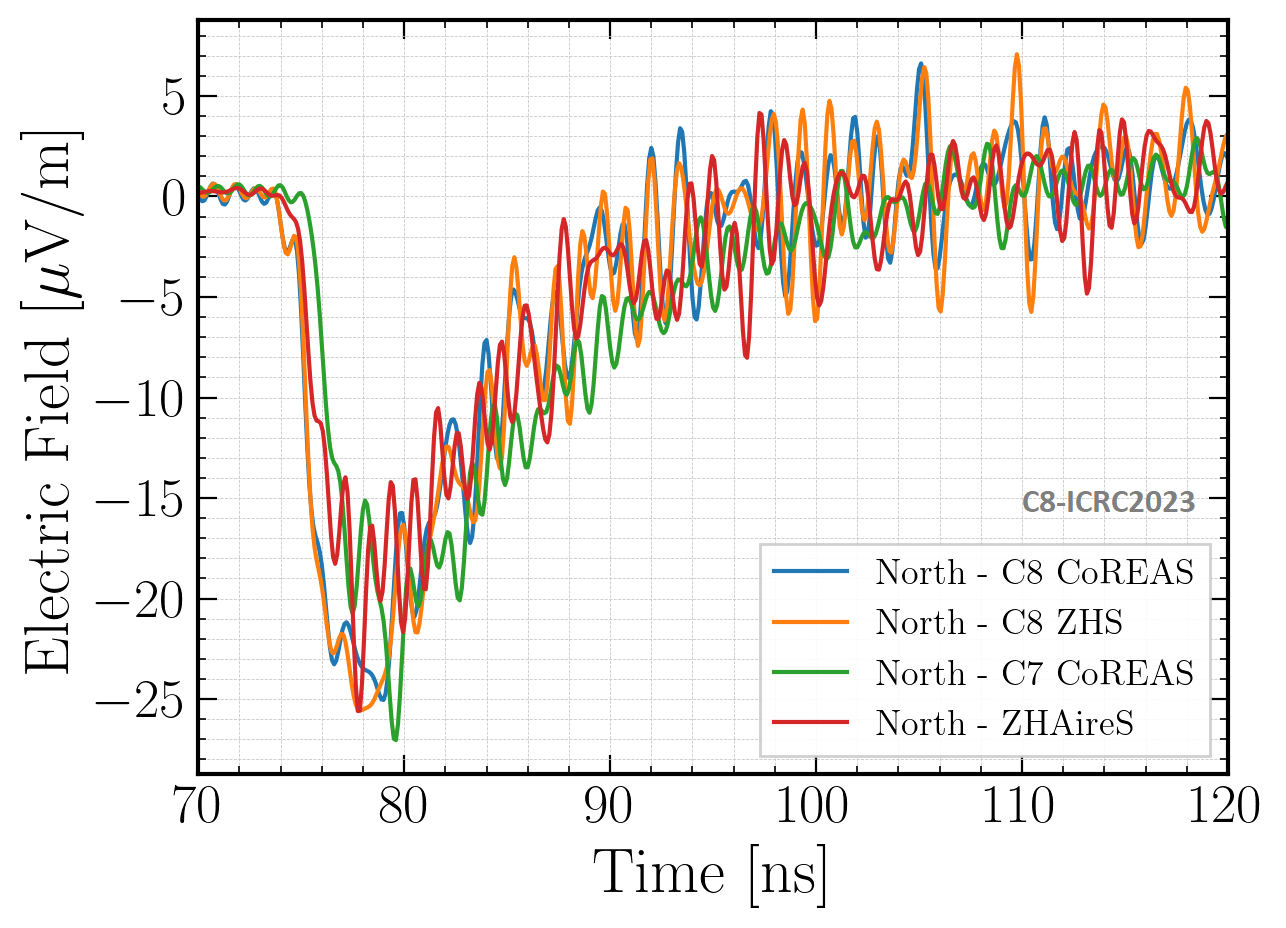}
    \subcaption[c]{\SI{200}{\metre}.}
    \label{fig:north200}
\end{subfigure} \hspace{10mm}
\begin{subfigure}{6cm}
    \centering
    \includegraphics[width=6cm]{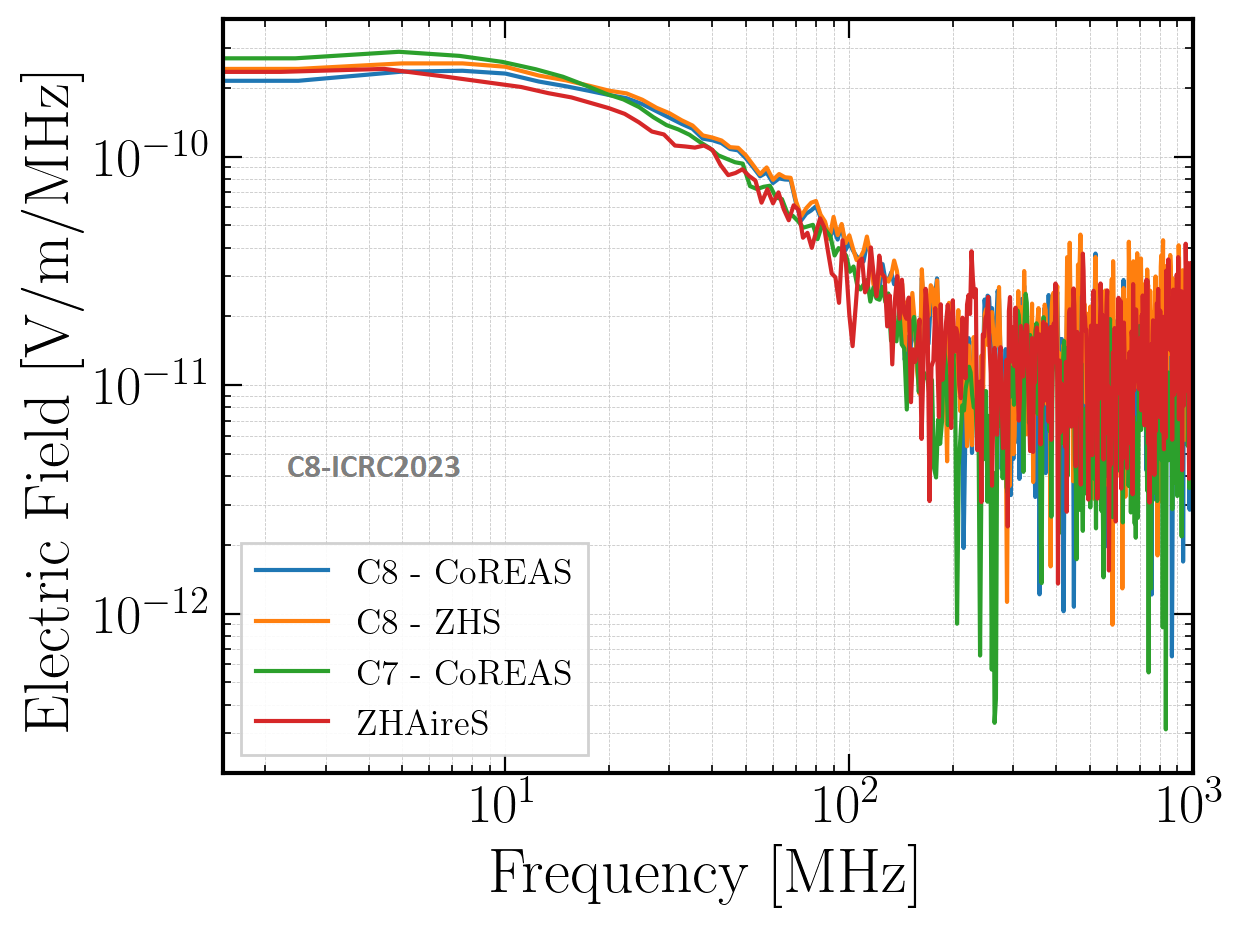}
    \subcaption[c]{\SI{200}{\metre}.}
    \label{fig:north200fft}
\end{subfigure}
\caption{Signal pulse and frequency spectra comparison in \SIrange{0}{1}{\giga\hertz} band, for various antenna distances from the shower core - Charge excess contribution.}
\end{figure}

\begin{figure}
    \centering
    \includegraphics[width=0.8\textwidth]{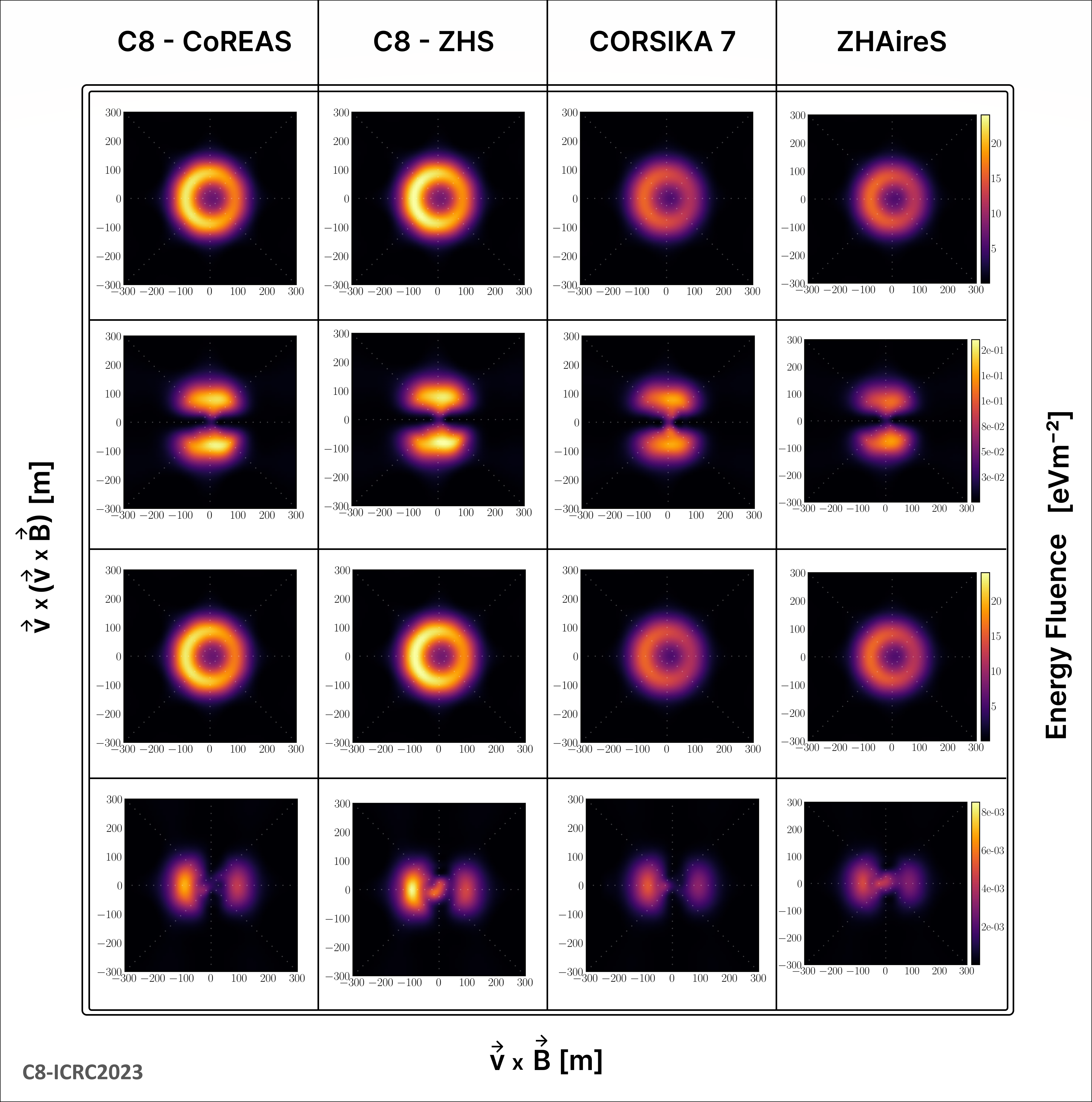}
    \caption{Table of energy fluence in different polarizations of the electric field in the \SIrange{50}{350}{\mega\hertz} frequency band for C8 CoREAS, C8 ZHS, C7 CoREAS and ZHAireS. The order of the polarizations we see starting from top to bottom is: all polarizations, $ \vec{v} \times (\vec{v} \times \vec{B}) $, $\vec{v} \times \vec{B}$ and $\vec{v}$.}
    \label{fig:maps}
\end{figure}

In this section we proceed to simulating the radio emission from a \SI{100}{\peta\eV} iron induced extensive air shower with the 3 air shower codes at our disposal, namely C8, C7 and ZHAireS. For our simulations we use the ``US Standard atmosphere'', the refractive index profile according to the Gladstone-Dale law with the refractive index at sea level set to $n=\num{1.000327}$, and a constant horizontal geomagnetic field of \SI{50}{\micro\tesla} aligned in the $x$ direction. We also use thinning for all the showers to make computing times better. The thinning implementations in the 3 shower codes differ though, we tried to set the parameters in a way so that thinning behaves in a very similar way across our simulations, in order to make our showers comparable. We used the same hadronic interaction model (Sibyll 2.3d~\cite{Riehn:2019jet}) and for C8 for the electromagnetic interaction model, we used PROPOSAL v7.6.1~\cite{Alameddine:2020zyd,proposal-zenodo}. It is worth mentioning that C8 also supports FLUKA~\cite{Battistoni:2015epi} now, which was used for the simulations. The electromagnetic particle energy cuts were set to \SI{0.5}{\MeV}, while hadronic and muonic energy cuts were set to \SI{0.3}{\giga\eV}. For the antenna array we use a star-shaped pattern of 160 antennas (sampling period is set to \SI{0.1}{\nano\second}) located at the ground, in 20 concentric rings spaced equally from \SIrange{25}{500}{\metre} from the shower axis with 8 antennas distributed azimuthally in each ring (exact locations are shown in the 2D fluence maps in \cref{fig:maps}).

The resulting showers can be seen in the longitudinal profiles (\cref{fig:profile}), where the number of electrons and positrons are plotted with respect to grammage. The showers are similar in terms of number of particles and \Xmax, where for C8 \Xmax is $\sim\SI{540}{\gram \per \cm \squared}$, for C7 $\sim\SI{570}{\gram \per \cm \squared}$ and for ZHAireS $\sim\SI{550}{\gram \per \cm \squared}$, so we go ahead and compare the radio emission from these showers.

When comparing the simulations from C8, C7 and ZHAireS one must take into account that each shower code uses a different tracking and thinning algorithms and different electromagnetic interaction models so a 100\% agreement is unrealistic since the showers are not identical. The structure of our comparison is as follows: we pick the arm of antennas in the $\vec{v} \times (\vec{v} \times \vec{B})$ direction, at $90^{o}$ from the positive x-axis and we compare filtered pulses in the \SIrange{0}{1}{\giga\hertz} frequency band range along with the frequency spectra in the same band for different radial distances from the shower core. We study them in polarizations "$\vec{v} \times \vec{B}$" (West) and " $ \vec{v} \times (\vec{v} \times \vec{B}) $" (North) in order to decouple and investigate separately the \textit{geomagnetic} and \textit{charge excess} contributions. Then, we restrict our study in a more realistic use case scenario and we plot the 2D fluence maps in the \SIrange{50}{350}{\MHz} band. 2D fluence maps in the narrower \SIrange{30}{80}{\MHz} band are also presented in~\cite{Tim}.

We first investigate the geomagnetic contribution and the results are promising. For an antenna at \SI{50}{\metre} the shower core (\cref{fig:west50}, \cref{fig:west50fft}) the difference in amplitude is within 5\% with C8 producing slightly higher radiation in both formalisms. For an antenna close to the Cherenkov ring at \SI{100}{\metre} from the shower core (\cref{fig:west100}, \cref{fig:west100fft}) the C8 pulses have consistently more power that accounts for roughly 10\% and we start to notice a slight time offset between the 3 shower codes. For an antenna further away at \SI{200}{\metre}  (\cref{fig:west200}, \cref{fig:west200fft}) where the signal gets weaker the difference in amplitude gets bigger at roughly 15\% and the time offset is more prominent. Moving on to the charge excess contribution of the same antennas we have, at \SI{50}{\metre} from the shower core (\cref{fig:north50}, \cref{fig:north50fft}), close to the Cherenkov ring (\cref{fig:north100}, \cref{fig:north100fft}) and at \SI{200}{\metre} from the shower core (\cref{fig:north200}, \cref{fig:north200fft}). In this case the agreement in terms of amplitude is better (less than $\sim 10\%$) with the most prominent differences spotted in the Cherenkov ring where C8 pulses are stronger.

\begin{figure}[]
\centering
\begin{subfigure}{.49\textwidth}
    \centering
    \includegraphics[width=\plotwidth]{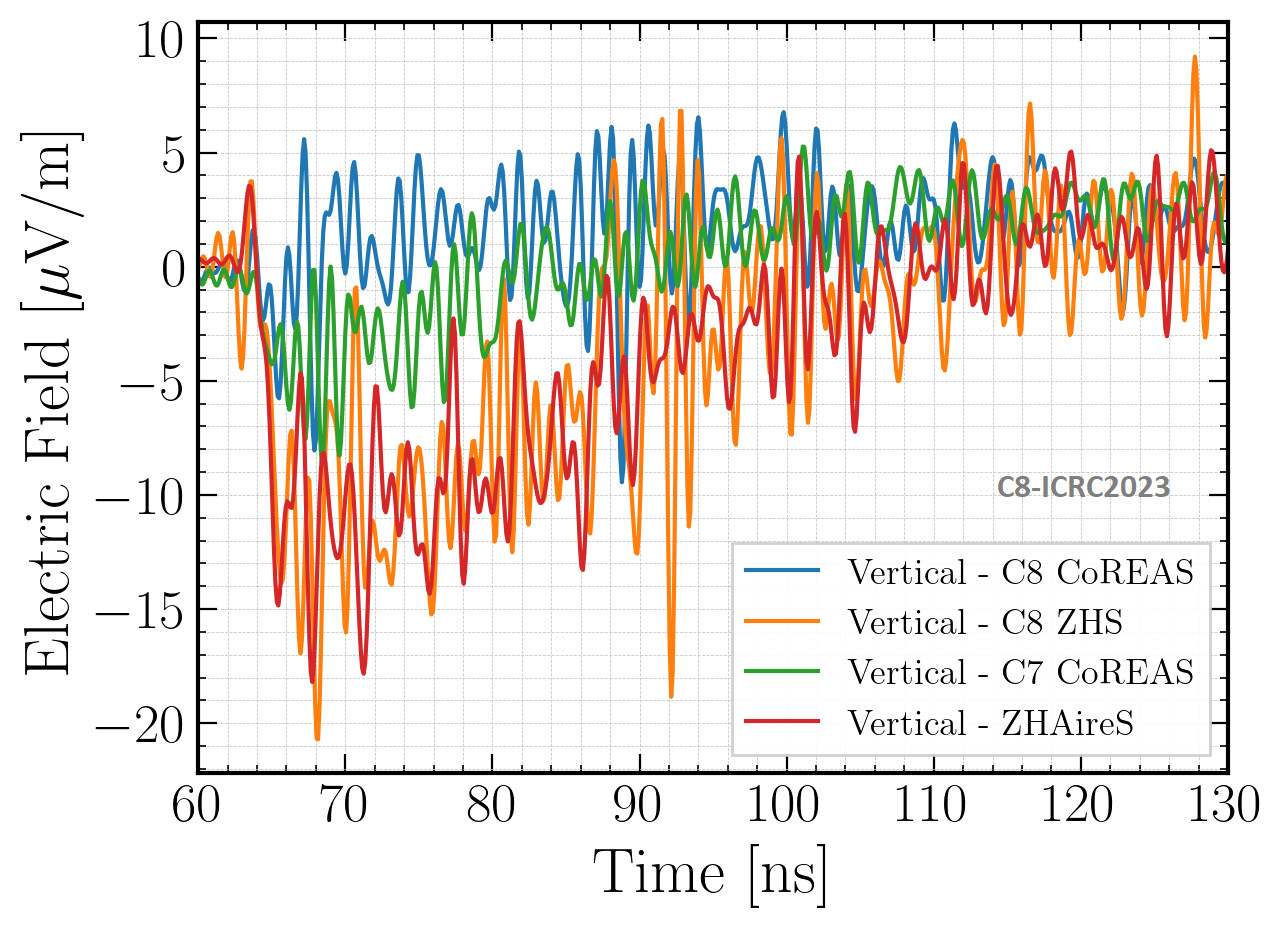}
\end{subfigure}
\begin{subfigure}{.49\textwidth}
    \centering
    \includegraphics[width=\plotwidth]{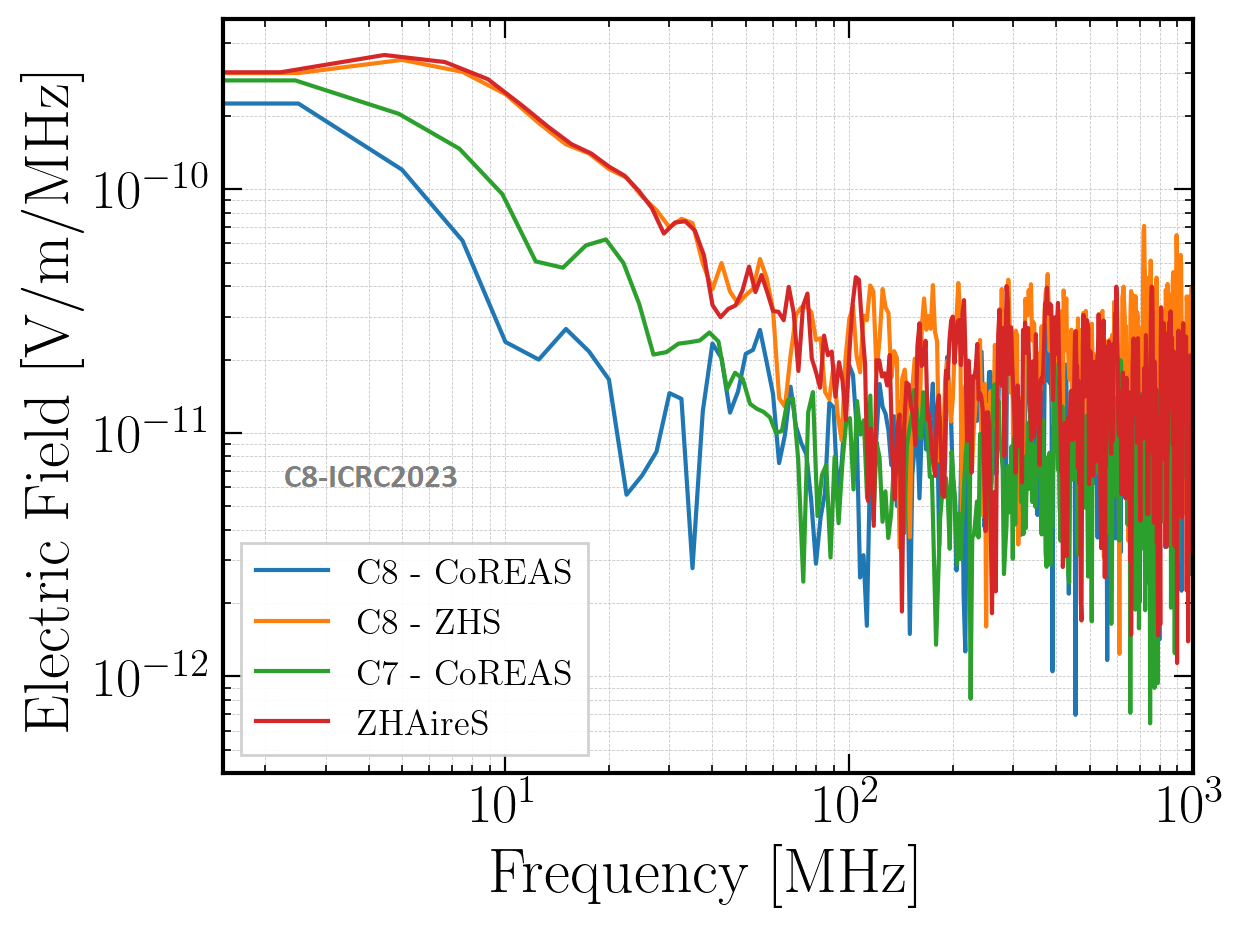}
\end{subfigure}
\caption{Signal pulse and frequency spectra comparison in \SIrange{0}{1}{\giga\hertz} band, for antenna at \SI{25}{\m} from the shower core - Vertical polarization.}
\label{fig:vert25}
\end{figure}

To examine the overall agreement between the shower codes we plot the 2D fluence map in the \SIrange{50}{350}{\MHz} frequency band (\cref{fig:maps}). Overall, in terms of energy deposit to the ground, footprint shape and symmetry the two formalisms simulated with C8 match C7 and ZHAireS quite well. We can confirm though, also in this plot that the C8 pulses contain roughly 10\% more power with respect to C7 and ZHAireS. It is worth pointing again at this point that we do not compare identical showers, so perfect agreement is not expected. The vertical polarization in the 2D fluence map shows an interesting feature. We can see that the two formalisms (CoREAS for C8 and C7 and ZHS for C8 and ZHAireS) produce a different prediction for the radio emission in the center and very close to the shower core, as seen in the form of a "double blip". To confirm this, we plot the pulses and the frequency spectra for the vertical polarization for an antenna at \SI{25}{\metre} from the shower core (\cref{fig:vert25}), where we can see that C8-ZHS and ZHAireS predict $\sim50\%$ more power with respect to C8-CoREAS and C7 especially in frequencies below \SI{500}{\mega\hertz}. A similar behavior is seen also in the \SIrange{30}{80}{\mega\hertz} 2D fluence maps~\cite{Tim}, as expected. In general, the vertical polarization is not practically relevant though, since it is very small when compared to the dominant $\vec{v} \times \vec{B}$ fluence.

 \section{Summary}
 
We have presented the advancements and improvements of the radio module in C8 since ICRC~2021 and ARENA~2022 along with a full scale comparison of the radio emission from a \SI{100}{\peta\eV} iron-induced extensive air shower. We have reached a very good agreement with CORSIKA~7 and ZHAireS. CORSIKA~8 as a whole has made significant progress to reach a stage where we trust our simulations and CORSIKA~8 radio in particular, is ready for potential use.

\printbibliography

\clearpage

\section*{The CORSIKA 8 Collaboration}
\small

\begin{sloppypar}\noindent
\input{latex_authorlist_authors}
\end{sloppypar}

\begin{center}
\rule{0.1\columnwidth}{0.5pt}
\raisebox{-0.4ex}{\scriptsize$\bullet$}
\rule{0.1\columnwidth}{0.5pt}
\end{center}

\vspace{-1ex}
\footnotesize
\input{latex_authorlist_institutions}

\vspace{-1ex}
\footnotesize
\input{acknowledgments}



%
%
%

\end{document}

%% file: latex_authorlist_authors.tex
J.M.~Alameddine$^{1}$,
J.~Albrecht$^{1}$,
J.~Alvarez-Mu\~niz$^{2}$,
J.~Ammerman-Yebra$^{2}$,
L.~Arrabito$^{3}$,
J.~Augscheller$^{4}$,
A.A.~Alves Jr.$^{4}$,
D.~Baack$^{1}$,
K.~Bernl\"ohr$^{5}$,
M.~Bleicher$^{6}$,
A.~Coleman$^{7}$,
H.~Dembinski$^{1}$,
D.~Els\"asser$^{1}$,
R.~Engel$^{4}$,
A.~Ferrari$^{4}$,
C.~Gaudu$^{8}$,
C.~Glaser$^{7}$,
D.~Heck$^{4}$,
F.~Hu$^{9}$,
T.~Huege$^{4,10}$,
K.H.~Kampert$^{8}$,
N.~Karastathis$^{4}$,
U.A.~Latif$^{11}$,
H.~Mei$^{12}$,
L.~Nellen$^{13}$,
T.~Pierog$^{4}$,
R.~Prechelt$^{14}$,
M.~Reininghaus$^{15}$,
W.~Rhode$^{1}$,
F.~Riehn$^{16,2}$,
M.~Sackel$^{1}$,
P.~Sala$^{17}$,
P.~Sampathkumar$^{4}$,
A.~Sandrock$^{8}$,
J.~Soedingrekso$^{1}$,
R.~Ulrich$^{4}$,
D.~Xu$^{12}$,
E.~Zas$^{2}$

%% file: latex_authorlist_institutions.tex
\begin{description}[labelsep=0.2em,align=right,labelwidth=0.7em,labelindent=0em,leftmargin=2em,noitemsep]
\item[$^{1}$] Technische Universit\"at Dortmund (TU), Department of Physics, Dortmund, Germany
\item[$^{2}$] Universidade de Santiago de Compostela, Instituto Galego de F\'\i{}sica de Altas Enerx\'\i{}as (IGFAE), Santiago de Compostela, Spain
\item[$^{3}$] Laboratoire Univers et Particules de Montpellier, Universit\'e de Montpellier, Montpellier, France
\item[$^{4}$] Karlsruhe Institute of Technology (KIT), Institute for Astroparticle Physics (IAP), Karlsruhe, Germany
\item[$^{5}$] Max Planck Institute for Nuclear Physics (MPIK), Heidelberg, Germany
\item[$^{6}$] Goethe-Universit\"at Frankfurt am Main, Institut f\"ur Theoretische Physik, Frankfurt am Main, Germany
\item[$^{7}$] Uppsala University, Department of Physics and Astronomy, Uppsala, Sweden
\item[$^{8}$] Bergische Universit\"at Wuppertal, Department of Physics, Wuppertal, Germany
\item[$^{9}$] Peking University (PKU), School of Physics, Beijing, China
\item[$^{10}$] Vrije Universiteit Brussel, Astrophysical Institute, Brussels, Belgium
\item[$^{11}$] Vrije Universiteit Brussel, Dienst ELEM, Inter-University Institute for High Energies (IIHE), Brussels, Belgium
\item[$^{12}$] Tsung-Dao Lee Institute (TDLI), Shanghai Jiao Tong University, Shanghai, China
\item[$^{13}$] Universidad Nacional Aut\'onoma de M\'exico (UNAM), Instituto de Ciencias Nucleares, M\'exico, D.F., M\'exico
\item[$^{14}$] University of Hawai'i at Manoa, Department of Physics and Astronomy, Honolulu, USA
\item[$^{15}$] Karlsruhe Institute of Technology (KIT), Institute of Experimental Particle Physics (ETP), Karlsruhe, Germany
\item[$^{16}$] Laborat\'orio de Instrumenta\c{c}\~ao e F\'\i{}sica Experimental de Part\'\i{}culas (LIP), Lisboa, Portugal
\item[$^{17}$] Fluka collaboration
\end{description}

%% file: acknowledgments.tex
\section*{Acknowledgments}
This research was funded by the Deutsche Forschungsgemeinschaft (DFG, German Research Foundation) – Projektnummer 445154105. For the simulations presented, computing resources from KIT have been used. This work has also received financial support from Xunta de Galicia (Centro singular de investigación de Galicia accreditation 2019-2022), by European Union ERDF, by the "María de Maeztu" Units of Excellence program MDM-2016-0692, the Spanish Research State Agency and from Ministerio de Ciencia e Innovación PID2019-105544GB-I00 and RED2018-102661-T (RENATA).